\documentclass[pra,aps,showpacs,twocolumn,floatfix]{revtex4}
\usepackage{graphicx}
\usepackage{array}
\usepackage{color}
\usepackage{amsmath}
\usepackage{amsxtra}
\usepackage{amstext}
\usepackage{amssymb}
\usepackage{latexsym}
\usepackage{dsfont}
\usepackage{cases}
\usepackage{mathrsfs}
\usepackage{color}
\usepackage{times}
\usepackage{natbib}
\begin{document}

\bibliographystyle{plainnat}

\title{Implementing conventional and unconventional nonadiabatic geometric quantum gates via SU(2) transformations}
\author{Jian-jian Cheng}
\author{Lin Zhang}\email{zhanglincn@snnu.edu.cn}
\affiliation{School of physics and information technology, Shaanxi Normal University,Xi'an 710119, P. R. China}
\begin{abstract}
We propose a simple but versatile protocol to engineer time-dependent Hamiltonians inversely for geometric quantum
computation. By utilizing SU(2) transformation, a speedup goal on gate operation is achieved
with more freedom to design the control parameters. As an application, this protocol enables the conventional
and unconventional nonadiabatic geometric quantum gates with desired evolution paths by controlling the microwave
pulses in the diamond nitrogen-vacancy center system. We show that the inversely designed Hamiltonian can fulfill the geometric
gate with more economical evolution time and further reduces the influence of the environment noise on gate fidelity.
\end{abstract}

%\pacs{03.65.Ca, 03.65.Xp, 03.65.Vf}
\maketitle

\section{Introduction}

Quantum computation exhibits more efficiency than the classical one in solving some problems,
such as factoring large integers, searching big databases, and finding optimal solutions by quantum
annealing \cite{Shor,Grover,Johnson}. However, it still faces enormous challenges in theory and
applications mainly due to the inevitable noises introduced by the control errors and the
system-environment couplings, which destroy the state coherence that plays a kernel role in
the parallel computation.

To overcome this challenge, geometric quantum computation (GQC) has been proposed \cite{Zanardi}.
As the geometric phase is solely related to the structure of an evolution path and independent of the middle
details, a quantum gate designed with geometric phase is immune to local disturbances during the evolution \cite{ev1,ev2,ev3}.
However, the scheme of the geometric gate based on adiabatic Abelian or non-Abelian geometric phase \cite{berry,AA}
suffers slowly cyclic evolutions. Although the adiabatic geometric gates are robust against control errors,
the lengthy gate operation time for holonomic quantum computation is still vulnerable to the environment-induced
decoherence \cite{zong}. To relax the limit of evolution speed, non-adiabatic holonomic quantum computation (NHQC)
based on nonadiabatic non-Abelian geometric phase was proposed by constructing driving Hamiltonians with time-independent
eigenstates \cite{NH1,NH2,NH3,NH4,NH5,NH6,NH7,NH8,NH9,NH10,NH11,NH12,NH13}. It has been proved that the implementation of
high-speed gates for quantum computation is plausible \cite{Long}.

As the traditional GQC should undergo cyclic evolutions and eliminate the dynamical phase to keep its gauge invariance,
the evolution paths were mainly restricted to special forms such as the former multiple loops and the newly
orange-slice-shaped loops \cite{O1,O2,O3,O4,O5}. The multiple-loop scheme adopts several closed loops to cancel out
the dynamical phase and the orange-slice-shaped-loop takes the geodesic path on Bloch sphere to eliminate the
dynamical phase during the evolution. Although the paths in the orange-slice-shaped-loop scheme are generally
shorter than those in the multiple-loop scheme, they still take a longer time to realize geometric computation
beyond decoherence time and do not integrate well with the experiments. Hence, how to optimize the evolutionary
paths for realizing nonadiabatic geometric quantum computation becomes a topic with great interest.

In this paper, we propose a reverse design scheme for GQC in two-level systems. By using a universal SU(2)
transformation to design the evolutionary path of the system, a speedup goal of the geometric gate can be
achieved in principle with more freedom to select the control parameters. Due to the flexibility of this approach,
the conventional and unconventional nonadiabatic geometric gates with desired evolution paths can be
freely designed under the conditions of cyclic evolution and parallel transportation. This approach is much more
powerful to find better evolutionary paths and can be well integrated with the experiments. As a demonstration,
we adopt the diamond nitrogen-vacancy (NV) center system as a platform to illustrate this approach. The nonadiabatic
geometric gate in this system can be realized by manipulating solid-state spins in the NV center by appropriately
controlling the amplitude, phase, and frequency of the pulsing fields. Our approach allows us to execute geometric gates
with shorter operation times than those in the previous schemes, so the influence of environment noises on the quantum
gates can reduce further. In particular, our inverse-engineering framework is more compatible and conducive to solving
the parametric matching problem for different computing platforms.

\section{GENERAL FRAMEWORK BY TRANSFORMATION Method}
In this section, we will give a general idea of the transformation method to realize feasible nonadiabatic
geometric quantum computation. Consider a two-level (one-qubit) quantum system described
by $(\hbar=1)$
\begin{equation}%
\hat{H}_{0}(t)=h_{x}(t)\hat{\sigma}_{x}+h_{y}(t)\hat{\sigma}_{y}+h_{z}(t)\hat{\sigma}_{z},\label{H0}
\end{equation}
where $h_{k}(t)(k=x,y,z)$ are arbitrary real functions of a time to be designed, and $\hat{\sigma}_{x},%
\hat{\sigma}_{y}, \hat{\sigma}_{z}$ are Pauli operators. The state evolution of the system is given by
\begin{equation}%
|\Psi(t)\rangle=\hat{U}_0(t)|\Psi(0)\rangle,
\end{equation}
where the initial state $|\Psi(0)\rangle$ is reset by initialization and the evolution operator $\hat{U}_0(t)$
will be properly designed. To engineer feasible Hamiltonians $H_{0}(t)$ that give desired dynamics
$\hat{U}_0(t)$, we adopt a unitary transformation of $R(t)$ parameterized by \cite{Unit}
\begin{equation*}
R\left( t\right) \equiv R\left( \theta ,\varphi \right) =\left[
\begin{array}{cc}
\cos \frac{\theta \left( t\right) }{2} & -e^{-i\varphi \left( t\right) }\sin
\frac{\theta \left( t\right) }{2} \\
e^{i\varphi \left( t\right) }\sin \frac{\theta \left( t\right) }{2} & \cos
\frac{\theta \left( t\right) }{2}%
\end{array}%
\right] ,
\end{equation*}
which enables a representation transformation of $|\psi(t)\rangle=R^{\dagger}|\Psi(t)\rangle$. The corresponding
Sch\"{o}rdinger equation is \cite{L.Zhang}
\begin{equation}%
i\frac{\partial}{\partial t}|\psi(t)\rangle = \hat{H}_R(t)|\psi(t)\rangle,
\end{equation}
and the transformed Hamiltonian is
\begin{equation}%
\hat{H}_R(t)=R^{\dagger}\hat{H}_{0}(t)R+i\partial_{t}R^{\dagger}R,\label{HRt}
\end{equation}
where $R^{\dagger}\hat{H}_{0}(t)R$ is often called the dynamical part, which is related to dynamical phase and
$i\partial_{t}R^{\dagger}R$ is the non-Abelian part, which often brings geometric phase  \cite{non}. Then
the time-evolution operator in $R$ representation reads
\begin{equation}\label{URt}
\hat{U}_{R}(t)=\mathcal{\hat{T}} \exp\left[-i\int_{0}^{t}\hat{H}_{R}(t^{\prime})dt^{\prime}\right].
\end{equation}

To remove the time ordering operator $\mathcal{\hat{T}}$ to calculate $\hat{U}_R(t)$, we can design
a diagonal form of
\begin{equation}\label{diag}
\hat{H}_R(t)=F(t)\hat{\sigma}_{z}
\end{equation}
by opportunely choosing the transformation parameters $\theta(t)$ and $\varphi(t)$ in $R(t)$. In this case,
the time evolution operator becomes
\begin{equation}
\hat{U}_{R}(t)=e^{-i\int_{0}^{t}\hat{H}_R(t^{\prime})dt^{\prime}}=e^{-i\gamma(t)\hat{\sigma}_{z}},
\end{equation}
where $\gamma(t)=\int_{0}^{t}F(t^{\prime})dt^{\prime}$. Therefore, the time-evolution operator in the previous
representation can be obtained by
\begin{equation}%
\label{U0t}
\hat{U}_{0}(t)=R(t)\hat{U}_{R}(t)R^{\dagger}(0).
\end{equation}

To design $\hat{H}_R(t)$ to retain only the diagonal part, a good choice is to make the
non-diagonal parts of $R^{\dagger}\hat{H}_{0}(t)R$ and $i\partial_{t}R^{\dagger}R$ cancel out at
any time. Then the diagonal matrices $K(t)\equiv \mathrm{dig} [R^{\dagger}H_{0}(t)R]$ and
$A(t)\equiv \mathrm{dig} [i\partial_{t}R^{\dagger}R]$ will safely lead to Eq.(\ref{diag}).
This diagonalization procedure is equivalent to set a parallel transportation condition for the quantum computation,
which means no transitions admit between two evolution states in $R$-representation. In order
to confine the control freedom for a reliable design, we consider a special
scheme that $K(t)$ is proportional to $A(t)$, i.e., $K(t)=\eta A(t)$ \cite{Wang2,Wang1}, which gives
\begin{equation}
\gamma(t)\hat{\sigma}_{z}=\int_{0}^{t}F(t^{\prime})\hat{\sigma}_{z}dt^{\prime}=\int_{0}^{t}(1+\eta)%
A(t^{\prime})dt^{\prime},\label{phase}
\end{equation}
where $\eta$ is a newly introduced constant parameter and $\eta\neq-1$ to avoid a trivial
case. The physical meaning of $\eta$ can be seen if we set $\eta=0$, the rotation phase, $\gamma(t)\hat{\sigma}_{z}\equiv\gamma_g(t)\hat{\sigma}_{z}=\int_{0}^{t}A(t^{\prime})dt^{\prime}$,
reduces to a pure geometric phase without any dynamical component. To discriminate them, we
denote the total phase $\gamma(t)$ as $\gamma(t)=(1+\eta) \gamma_g(t)$.

Another condition that needs to be satisfied in quantum geometric gate is the cyclic evolution,
which leads to recycling transformation $R(\tau)=R(0)$, i.e., $R(\theta_{\tau},\varphi_{\tau})= R(\theta_{0},\varphi_{0})$,
where the parameters are labeled by $\theta(t)\equiv \theta_{t}, \varphi(t)\equiv \varphi_{t}$ for convenience,
then the previous evolution operator will be
\begin{equation}
\label{U1t}
\hat{U}_0(\tau)=R(0)\hat{U}_{R}(\tau)R^{\dagger}(0)=e^{-i\gamma \mathbf{n}_0\cdot\hat{\boldsymbol{\sigma}}},
\end{equation}
where $\mathbf{n}_0\equiv(\sin\theta_{0}\cos\varphi_{0},\sin\theta_{0}\sin\varphi_{0},\cos\theta_{0})$
is a unit vector and  $\hat{U}_0(\tau)$ conducts a rotation around $\mathbf{n}_0$ by an angle $2\gamma(\tau)=2(1+\eta) \gamma_g(\tau)$.
Therefore, an arbitrary geometric gate for a single-qubit can be designed for the original system $\hat{H}_0(t)$.
In the spherical parametric space of $(1,\theta,\varphi)$,
the transition curve traces a closed path $C$ of a cyclic evolution during $t=0$ to $\tau$ and $\gamma_g(\tau)$ represents a half of
the solid angle enclosed by path $C$. This implies that the geometric phase $\gamma_g(\tau)$ is only determined
by the evolution path of the parameters $\theta(t)$ and $\varphi(t)$, and
independent of the evolutionary details such as changing rates
of the parameters. This demonstrates that the gate is robust against the control errors and depends only on the topological aspects
of the evolution path. As a matter of fact, $\gamma_g(\tau)$ is invariant as long as the area enclosed by the path does not change,
and the rotation axis depends only on the initial parameters $\theta_{0}$ and $\varphi_{0}$.

Although the nonadiabatic geometric gates are robust against the control errors that do not change the area
enclosed by the evolution path, an unavoidable challenge remains that the gates are still vulnerable to
environment-induced decoherence and sequentially impede the experimental implementation. Fortunately, a
nonadiabatic geometric gate with a determined area has many different evolutionary paths which can conditionally
explore. As some paths have shorter evolution times than others, the influences of environmental noises on the
quantum gates can reduce.

Based on the above strategy, we propose a scheme of nonadiabatic GQC with an unconventional geometric phase that
the dynamical phase does not need to be avoided \cite{Wang2,Wang1}. Although the total phase $\gamma(t)$ accumulated
in the designed gate operation contains the dynamical component, it still relies on global geometric features and
the corresponding gate is also a kind of geometric one in a general sense. The total phase is proportional to the
geometric phase and the corresponding proportional coefficient is constant, independent of (or at least some)
parameters of the qubit system \cite{Wang2}. Under the circumstances, the total phase possesses the same geometric
robustness as the geometric phase and can be directly used to realize geometric gate \cite{un}. When $\eta=0$ the
dynamical phase is removed, our scheme will recover the conventional nonadiabatic schemes.

\section{Designed Hamiltonian}

The main features of our scheme can be summarized as follows:
(1) According to Eq.(\ref{U0t}), the evolution operator $\hat{U}_{0}(t)$ of system $\hat{H}_0(t)$ with feasible
$h_{x,y,z}(t)$ is designed by properly choosing $R(t)$ and $\hat{U}_{R}(t)$. (2) By the diagonalization procedure,
the target $\hat{U}_{0}(t)$ satisfies the parallel transport condition. (3) By choosing the closed path in parametric
space, $R(t)$ meets the cyclic evolution condition. (4) The control functions $h_{x,y,z}(t)$ are then determined and
the inversely designed Hamiltonian $H_{0}(t)$ is achieved. In the following, we elucidate the details.

We adopt a unitary transformation $R(t)$ on the Hamiltonian Eq.(\ref{H0}) and obtain explicit forms of
$R^{\dagger}H_{0}(t)R$ and $i\partial_{t}R^{\dagger}R$ by
\begin{equation*}
R^{\dagger }\hat{H}_{0}(t)R=f_{x}\left( t \right) \hat{\sigma}%
_{x}+f_{y}\left( t\right) \hat{\sigma}_{y}+f_{z}\left(t \right) \hat{\sigma}_{z},
\end{equation*}%
where%
\begin{eqnarray*}
f_{x}\left( \theta ,\varphi \right)  &=&\left( \cos \theta \cos ^{2}\varphi
+\sin ^{2}\varphi \right) h_{x}\left( t\right)  \\
&&-\sin ^{2} \frac{\theta }{2} \sin  2\varphi
h_{y}\left( t\right)  \\
&&-\sin \theta \cos  \varphi h_{z}\left(
t\right) ,
\end{eqnarray*}%
\begin{eqnarray*}
f_{y}\left( \theta ,\varphi \right)  &=&-\sin ^{2} \frac{\theta }{2}%
 \sin  2\varphi h_{x}\left( t\right)  \\
&&+\left( \cos \theta \sin ^{2}\varphi +\cos ^{2}\varphi \right) h_{y}\left(
t\right)  \\
&&-\sin \theta \sin \varphi h_{z}\left(
t\right) ,
\end{eqnarray*}%
\begin{eqnarray*}
f_{z}\left( \theta ,\varphi \right)  &=&\sin \theta \cos
 \varphi h_{x}\left( t\right)  \\
&&+\sin  \theta \sin \varphi \cdot h_{y}\left(
t\right) +\cos \theta h_{z}\left( t\right),
\end{eqnarray*}%
and%
\begin{equation*}
i\partial _{t}R^{\dagger }R=g_{x}\left( t \right) \hat{\sigma}%
_{x}+g_{y}\left( t \right) \hat{\sigma}_{y}+g_{z}\left(t\right) \hat{\sigma}_{z},
\end{equation*}%
where%
\begin{equation*}
g_{x}\left( \theta ,\varphi \right) =\frac{\dot{\theta}}{2}\sin \varphi +%
\frac{\dot{\varphi}}{2} \sin \theta \cos \varphi,
\end{equation*}%
\begin{equation*}
g_{y}\left( \theta ,\varphi \right) =-\dot{\frac{\theta }{2}}\cos \varphi +%
\frac{\dot{\varphi}}{2} \sin \theta \sin \varphi,
\end{equation*}%
\begin{equation*}
g_{z}\left( \theta ,\varphi \right) =\frac{\dot{\varphi}}{2}\left( 1-\cos \theta \right).
\end{equation*}%

To diagonalize $\hat{H}_R(t)$ according to Eq.(\ref{HRt}), the non-diagonal components
of $R^{\dagger}\hat{H}_{0}(t)R$ and $i\partial_{t}R^{\dagger}R$ should cancel out leading to
$f_{x}(t)=-g_{x}(t)$ and $f_{y}(t)=-g_{y}(t)$.
The diagonal components should be proportional to each other: $f_{z}(t)=\eta g_{z}(t)$.
Then we can naturally arrive the familiar geometric phase \cite{geometric}
\begin{equation}%
\gamma_{g}(\tau)=\frac{1}{2}\int_{0}^{\tau}\left[1-\cos\theta(t)\right]\dot{\varphi}(t)dt.\label{GeoPh}
\end{equation}

By putting the above integral  on a closed path in the parametric space of $(1,\theta,\varphi)$, it will be
\begin{equation}
\label{Geometric}
\gamma_{g}=\frac{1}{2}\oint_{C}(1-\cos\theta)d\varphi
\end{equation}
with a cyclic condition along the closed path, and the total rotation phase in Eq.(\ref{U1t}) reads
$\gamma=(1+\eta)\gamma_{g}$,
which indeed possesses a global geometric feature beyond the changing rates of the control parameters.

Based on the above conditions, the general geometric gates governed by Eq.(\ref{H0}) can be realized
by opportunely choosing $h_{x}(t), h_{y}(t)$ and $h_{z}(t)$ as follows:
\begin{eqnarray*}
h_{x}\left( t\right)  &=&\frac{\dot{\varphi}}{2}%
\left[ \eta -\left( 1+\eta \right) \cos \theta \right] \sin
\theta \cos \varphi-\frac{\dot{\theta}}{2}\sin \varphi , \\
h_{y}\left( t\right)  &=&\frac{\dot{\varphi}}{2}%
\left[ \eta -\left(1+\eta \right) \cos \theta \right] \sin
\theta \sin \varphi+\frac{\dot{\theta}}{2}\cos\varphi , \\
h_{z}\left( t\right)  &=&\frac{\dot{\varphi}}{2}\left[\sin^2 \theta+2\eta \cos
\theta \sin ^{2}\frac{\theta }{2} \right],
\end{eqnarray*}
and this inversely designed Hamiltonian $\hat{H}_0(t)$ is equivalent to a general two-level model of
\begin{equation}
\hat{H}_{0}(t)=\frac{1}{2}\left[
\begin{array}{cc}
\Delta \left( t\right)  & \Omega _{R}(t)e^{-i\phi \left( t\right) } \\
\Omega _{R}(t)e^{i\phi \left( t\right) } & -\Delta \left( t\right)
\end{array}%
\right] ,  \label{H0t}
\end{equation}%
where
\begin{eqnarray*}
\Omega _{R}(t) &=&\sqrt{\dot{\theta}^{2}+\dot{\varphi}^{2}\sin
^{2}\theta \left[ \eta -\left( 1+\eta \right) \cos \theta \right] ^{2}}, \\
\phi \left( t\right)  &=&\arctan \left[ \frac{\left( \eta -\frac{1+\eta }{2}%
\sin 2\theta \right) \dot{\varphi}\sin \varphi +\dot{\theta}\cos \varphi }{%
\left( \eta -\frac{1+\eta }{2}\sin 2\theta \right) \dot{\varphi}\cos \varphi
-\dot{\theta}\sin \varphi }\right] , \\
\Delta \left( t\right)  &=&\dot{\varphi}\left[ \sin ^{2}\theta
+2\eta \cos \theta \sin ^{2}\frac{\theta }{2} \right] .
\end{eqnarray*}%
It is well-known that Eq.(\ref{H0t}) can fulfill on many physical platforms such as the
laser-driven atomic system, the quantum dot spin, or Josephson junction system \cite{Jc1,Jc2,Jc3,Jc4,Jc5}
as well as the diamond NV center controlled by tailoring the parameters of microwave
or laser fields \cite{NV1,NV2,NV3,NV4,NV5}. This method is generally beyond the adiabatic dynamics
without any confinement of slowly varying parameters.

\section{Gate Implementation}

To demonstrate our approach, we design geometric rotation gates in the diamond NV center system
by controlling the light pulses as an example \cite{NV}. A given geometric rotation can realize
by tailoring $\Delta(t)$ and $\Omega_R(t)$ of the microwave pulses along the desired evolution
path in the parametric space of $(1, \theta,\varphi)$. In the following, we take
$\hat{U}_{Z}(\tau)=e^{-i\pi\hat{\sigma}_{z}/2}$ as a sample target gate.
\begin{figure}[htpb]
  \centering
  \includegraphics[width=110pt]{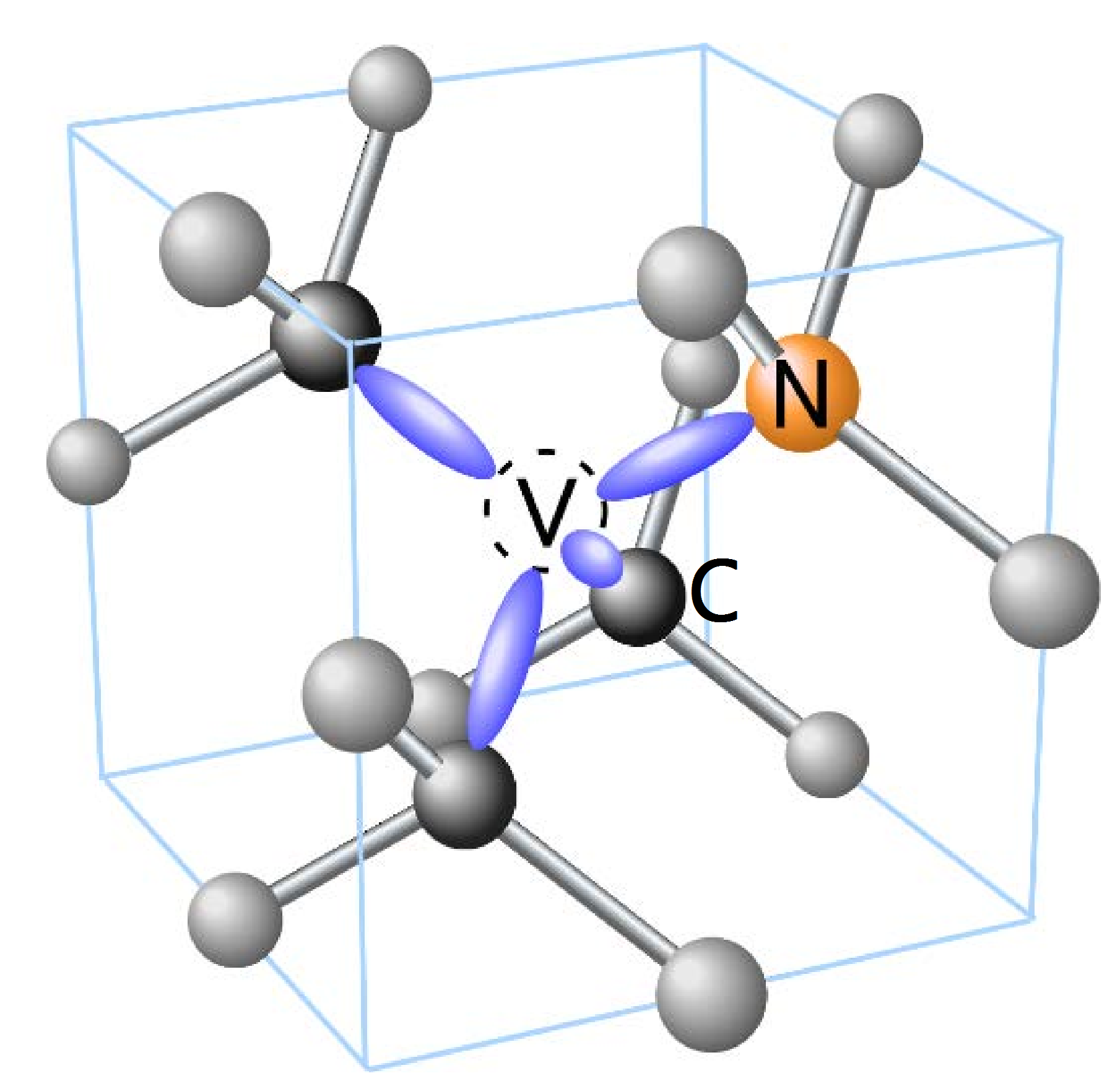}\,%
  \includegraphics[width=110pt]{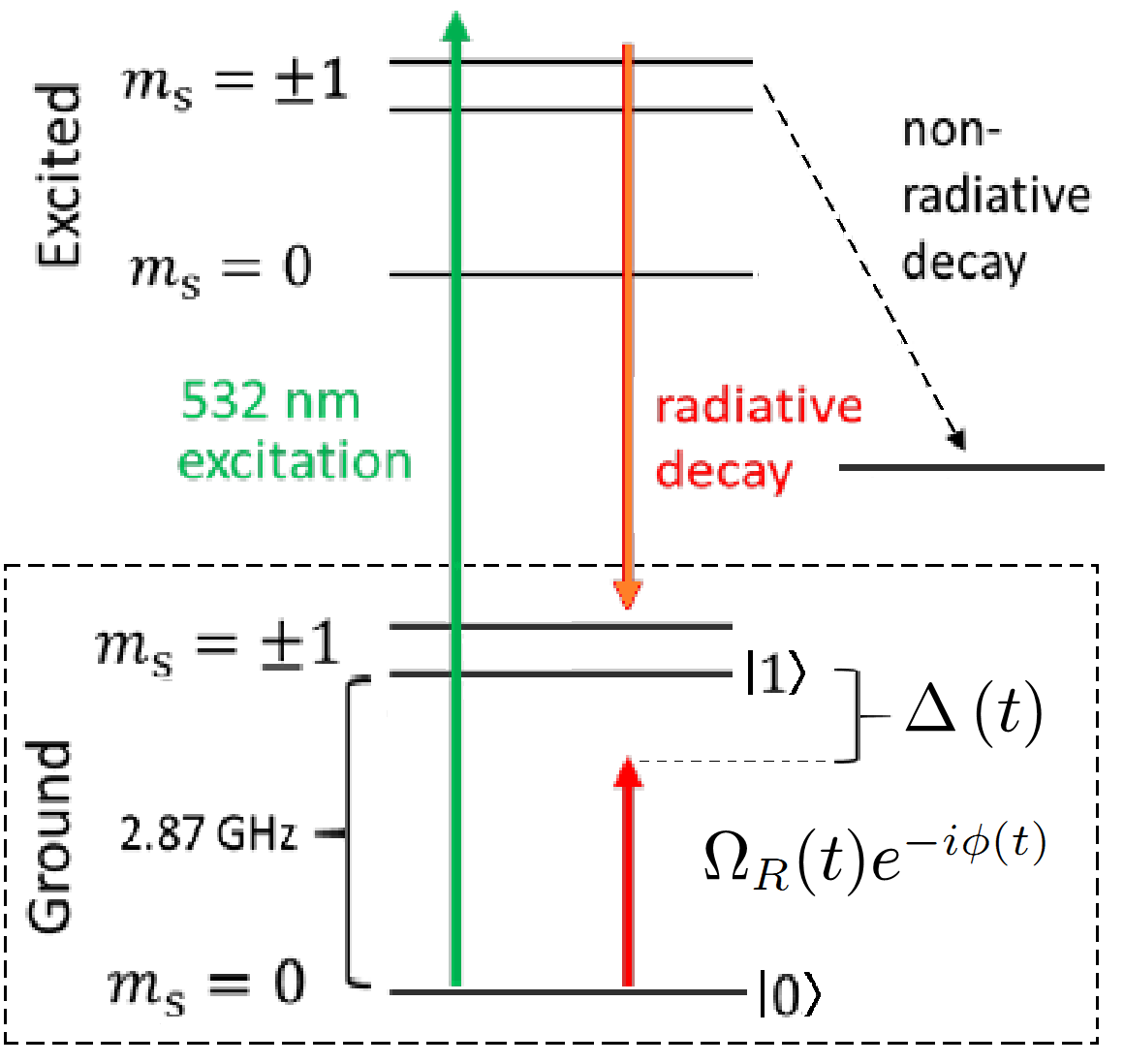}%
  \caption{Two Zeeman levels $|m_{s}=0\rangle$ and $|m_{s}=-1\rangle$ of the NV spin-triplet
ground state are encoded as the qubit states $|0\rangle$ and $|1\rangle$.}
  \label{figure1}
\end{figure}

As shown in Fig.\ref{figure1}, the NV center has a spin-triplet ground state and the nearby nuclear spins
($^{15}N$ and $^{13}C$) are polarized by a magnetic field of about $500\mathrm{G}$ along the NV axis. We
use two lower Zeeman levels $|m_{s}=0\rangle\equiv|0\rangle$ and $|m_{s}=-1\rangle\equiv|1\rangle$ of NV
center to encode the qubit and the nuclear spins of $^{13}C$ atom for further controls. The qubit is
manipulated by a microwave pulse whose spectrum, intensity and phase can be adjusted
by a hybrid waveform generator. The pulse parameters used here are $\Omega_{0}=20$ MHz
(the maximal Rabi frequency), $\Delta_0=20$ MHz (the maximal detuning) and the $\pi$-pulse control
time $\tau$ is in unit of $\tau_{0}=\pi/\Omega_{0}$ \cite{Duan}.

\subsection{Conventional geometric gate with ``orange-slice'' path}

\begin{figure}[htbp]
\begin{center}
\includegraphics[width=105pt]{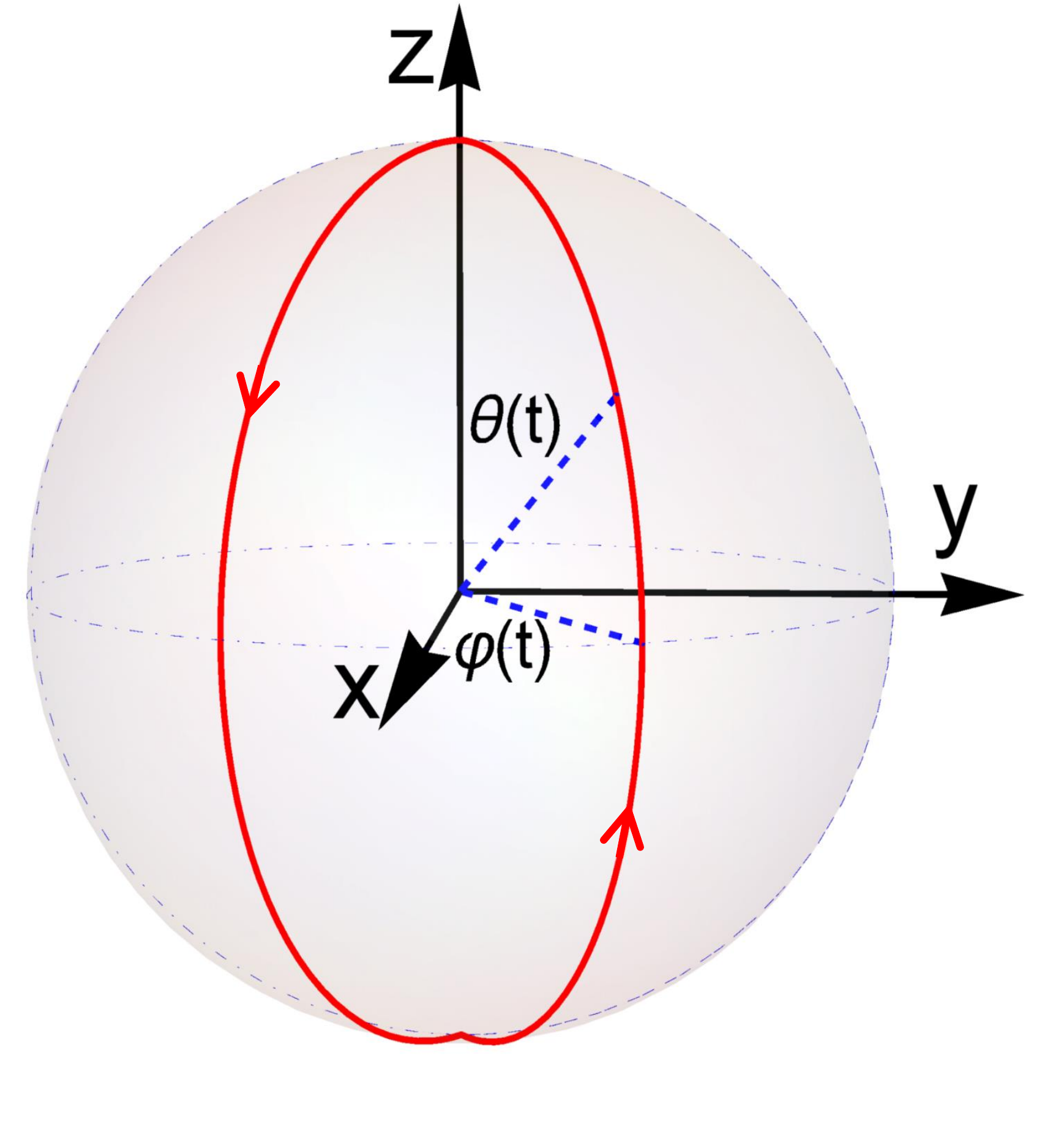}
\includegraphics[width=135pt]{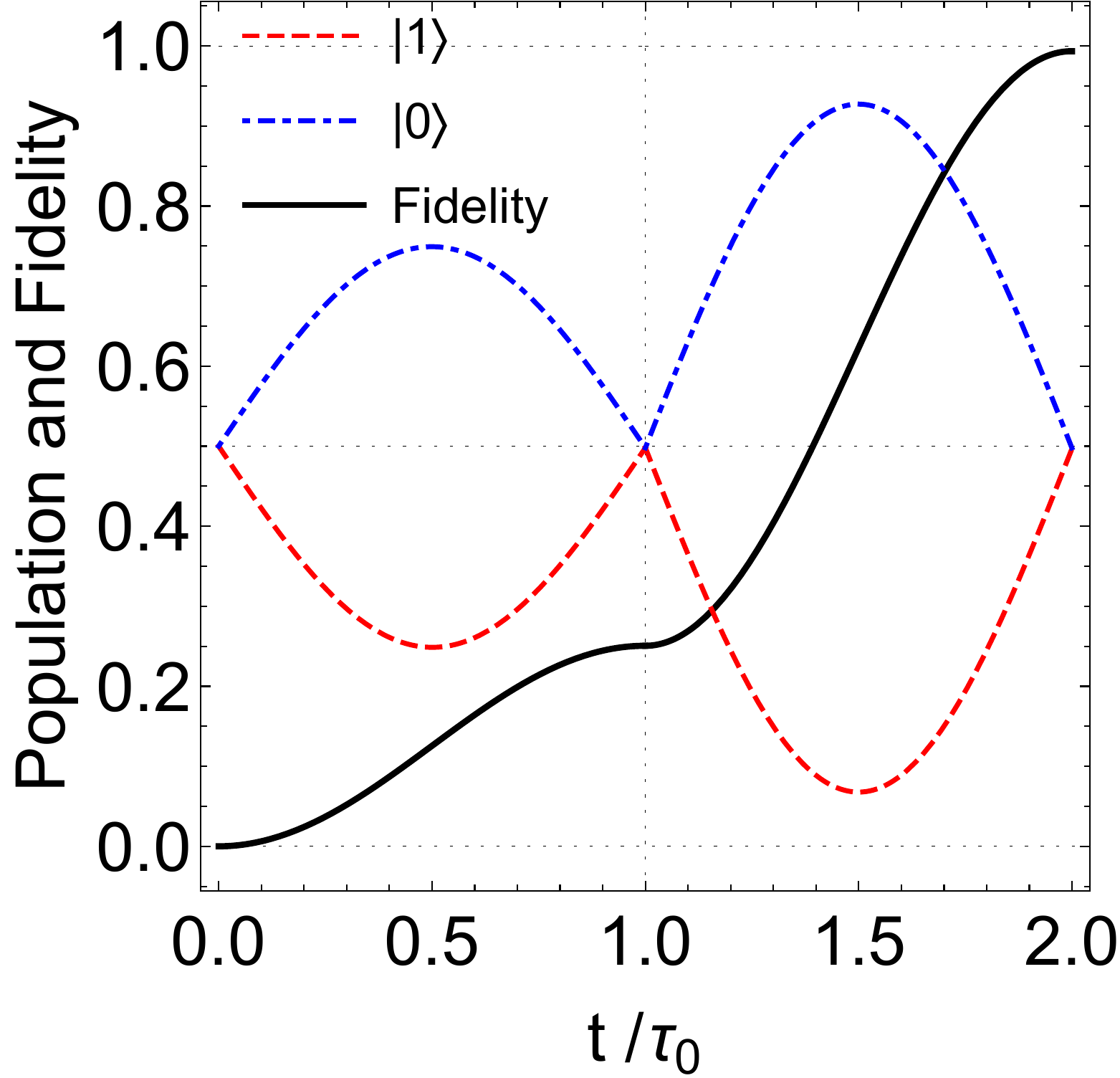}
\end{center}
\caption{(left) The orange-slice-shaped-loop path for the realization of $U_{Z}(\tau)$ gate;
(Right) The evolutions of state populations ($|1\rangle$: red dashed line; $|0\rangle$:
blue dot-dashed line) and the fidelities (black solid line) along two separate paths.}
\label{fig2}
\end{figure}

To realize gate $\hat{U}_{Z}(\tau)=e^{-i\frac{\pi}{2}\hat{\sigma}_{z}}$, a usual evolution path is selected
in parametric space as shown in Fig.\ref{fig2}. The parameters $[\theta(t),\varphi(t)]$ start from the north
pole $(0,\varphi_{0})$ to the south pole $(\pi,\varphi_{0})$ along the great circle $\varphi(t)=\varphi_{0}$,
then return back to the north pole from the south pole along another great circle $\varphi(t)=\varphi_{0}+\frac{\pi}{2}$.
This path is the so-called resonant ($\Delta=0$) orange-slice-shaped loop widely used in the control schemes
of nonadiabatic geometric quantum computation \cite{Tong2}.
To this end, the non-diagonal terms in Eq.(\ref{H0t}) are
%%%%
\begin{equation}
\label{Rabi1}
\Omega_{R}(t)e^{-i\phi(t)}=
\begin{cases}
\dot{\theta}(t)e^{-i(\varphi_{0}-\frac{\pi}{2})},\quad 0\leq t\leq\frac{\tau}{2},\\
\dot{\theta}(t)e^{-i\varphi_{0}},\quad \frac{\tau}{2}< t\leq \tau.
\end{cases}
\end{equation}
Here, the pulse areas of the Rabi frequencies at their respective time intervals satisfy
\begin{equation}
\label{Rabi2}
%\Omega_{R}(t)=
%\begin{cases}
\int_{0}^{\frac{\tau}{2}}\Omega_{R}(t)dt=\pi,\,
\int_{\frac{\tau}{2}}^{\tau}\Omega_{R}(t)dt=-\pi.
%\end{cases}
\end{equation}
If a square-wave pulse is used to do the calculation, the operation time is $\tau=2\tau_{0}=2\pi/\Omega_{0}$. The
geometric phase can be calculated from Eq.(\ref{Geometric}), $\gamma_{g}=\pi/2$,
which is obtained by the saltation of $\phi(t)$ at the moment of $t=\frac{\tau}{2}$ at the south pole.
That is how the conventional non-adiabatic geometric quantum gate is realized via the ``orange slice" as shown above. %in Fig.\ref{figure2}.

\subsection{Conventional geometric gate beyond ``orange-slice'' path}
\begin{figure}[htbp]
\begin{center}
\includegraphics[width=105pt]{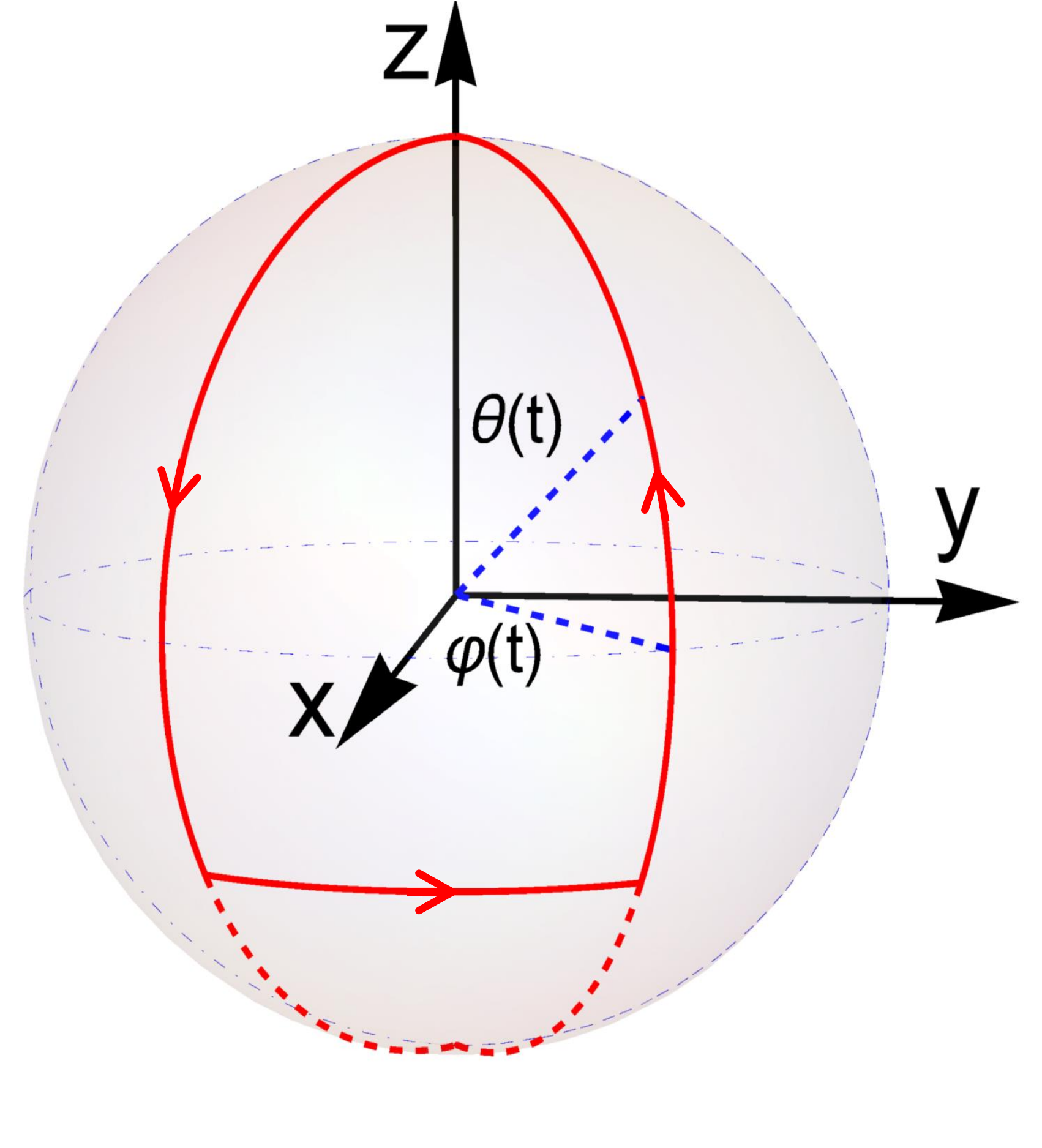}
\includegraphics[width=135pt]{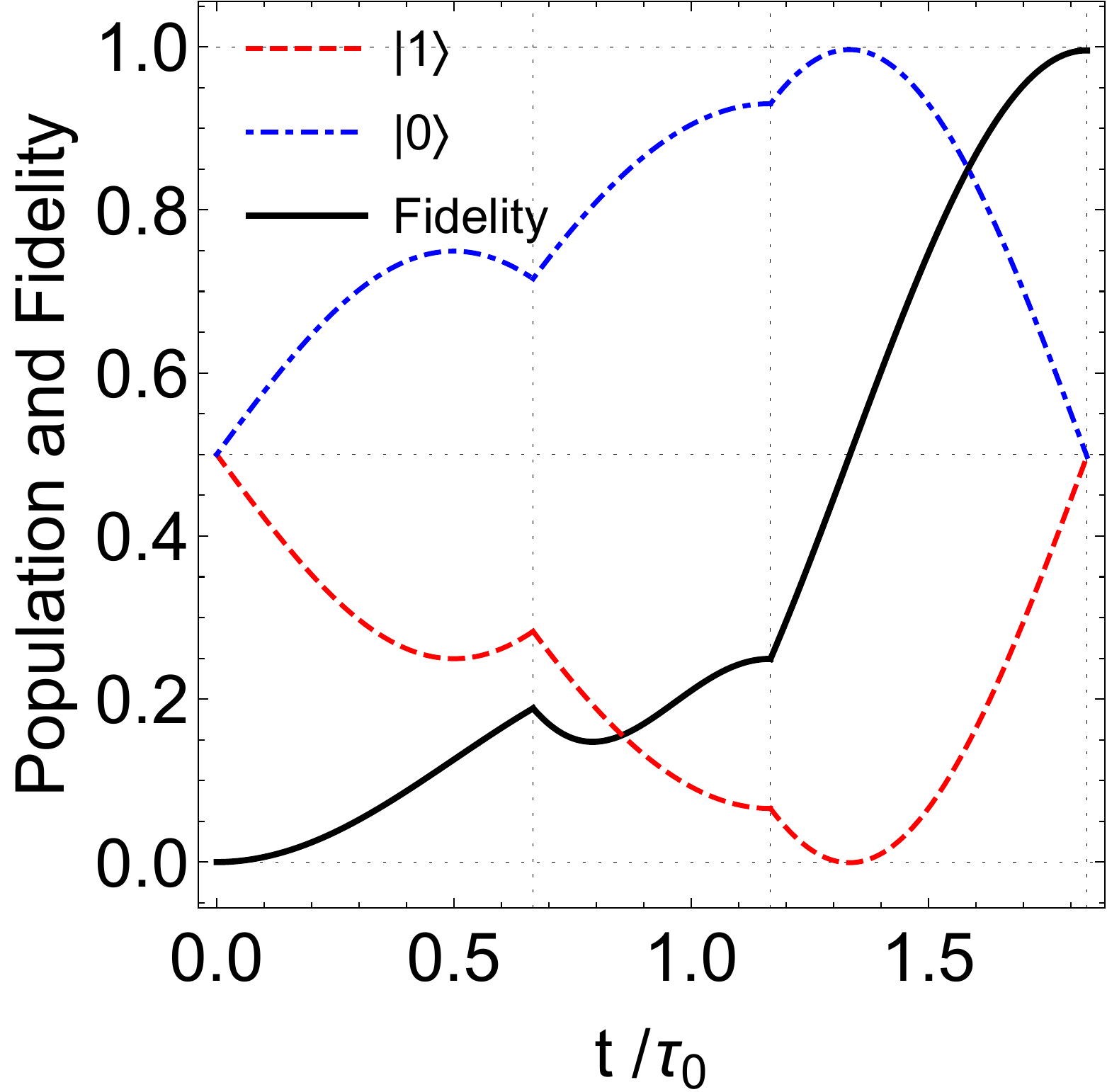}
\end{center}
\caption{(Left) The larger triangular path for the realization of $U_{Z}(\tau)$ gate;
(Right) The corresponding evolutions of state populations ($|1\rangle$: red dashed line; $|0\rangle$:
blue dot-dashed line) and the fidelities (black solid line) along three separate paths.}
\label{fig3}
\end{figure}

We can choose an alternative evolution path to realize $U_Z $ gate  to avoid the singular point at the
pole without saltation of $\phi$. As shown in Fig.\ref{fig3}, the parameters $[\theta(t),\varphi(t)]$
start from the north pole $(0,\varphi_{0})$ to the point $(\frac{2\pi}{3},\varphi_{0})$ along the great
circle $\varphi(t)=\varphi_{0}$. Then the parameters evolve from $(\frac{2\pi}{3},\varphi_{0})$ to
$(\frac{2\pi}{3},\varphi_{0}+\frac{2\pi}{3})$ along the arc $\theta(t)=\frac{2\pi}{3}$, and finally
return back to the north pole along the great circle $\varphi(t)=\varphi_{0}+\frac{2\pi}{3}$.

For this path, the Rabi frequencies of the laser pulses read
\begin{equation}
\label{Rabi}
\Omega_{R}(t)e^{-i\phi(t)}=
\begin{cases}
\dot{\theta}(t)e^{-i(\varphi_{0}-\frac{\pi}{2})},\, 0\leq t\leq\tau_{1},\\
\frac{\sqrt{3}}{4}\dot{\varphi}(t)e^{-i\varphi(t)},\,  \tau_{1}< t\leq\tau_{2},\\
\dot{\theta}(t)e^{-i(\varphi_{0}+\frac{\pi}{6})},\, \tau_{2}<t\leq\tau.
\end{cases}
\end{equation}
Here, their respective pulse areas and detunings are
\begin{eqnarray}
\label{Rabi2}
\int_{0}^{\tau_{1}}\Omega_{R}(t)dt &=&\frac{2\pi}{3},\,\Delta(t)=0,\notag\\
\int_{\tau_{1}}^{\tau_{2}}\Omega_{R}(t)dt &=&\frac{\sqrt{3}\pi}{6},\,\Delta(t)=\frac{3}{4}\dot{\varphi}(t), \\
\int_{\tau_{2}}^{\tau}\Omega_{R}(t)dt &=&-\frac{2\pi}{3},\,\Delta(t)=0.\notag
\end{eqnarray}
The total evolution time of this triangular path by the square-wave pulse is
$\frac{2}{3}\tau_0+\frac{1}{2}\tau_0+\frac{2}{3}\tau_0 \approx 1.833\tau_{0}$, which
is shorter than that of the orange-slice-shaped loop. Further, a shorter evolution path
to realize this gate can also be designed by this method if the Rabi frequency $\Omega(t)$
and detuning $\Delta(t)$ can reach the experimental maximum at the same time, and more optimal
evolution time will be about $1.792\tau_{0}$.

\subsection{Unconventional geometric gate}
Although the above triangular path evolves faster than the orange-slice one in the conventional geometric gate,
the requirement of zero dynamical phases imposes stringent constraints on the driving Hamiltonian.
However, our method relaxes the experimental conditions and combines geometric phase control with the non-adiabatic
dynamics to validate the dynamical phase ($\eta \neq 0$) in the gate design. Therefore, we can provide better
evolution paths with more relaxed experimental conditions to design $\Omega(t)$ and $\Delta(t)$.

\begin{figure}[htbp]
\begin{center}
\includegraphics[width=105pt]{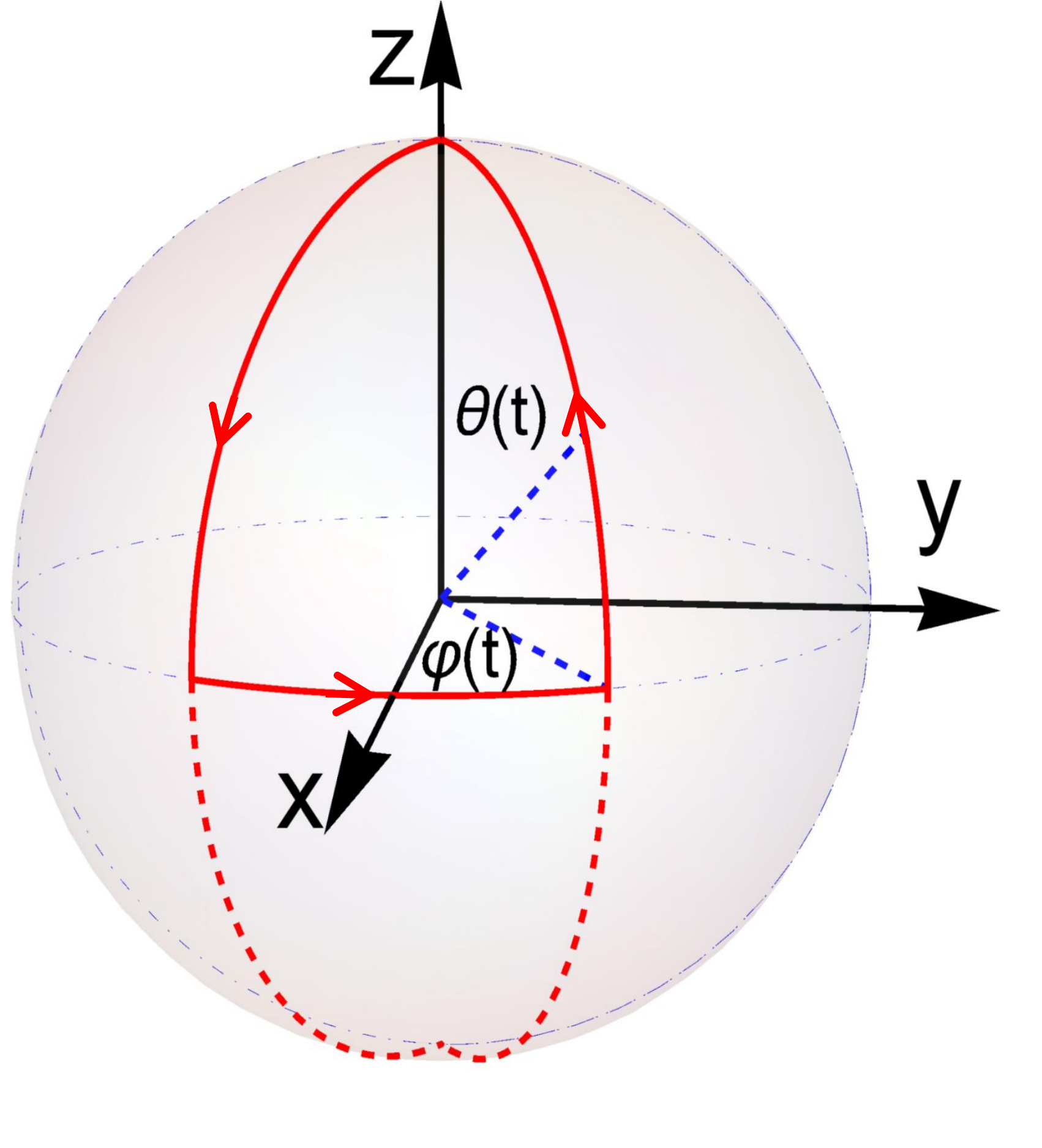}
\includegraphics[width=135pt]{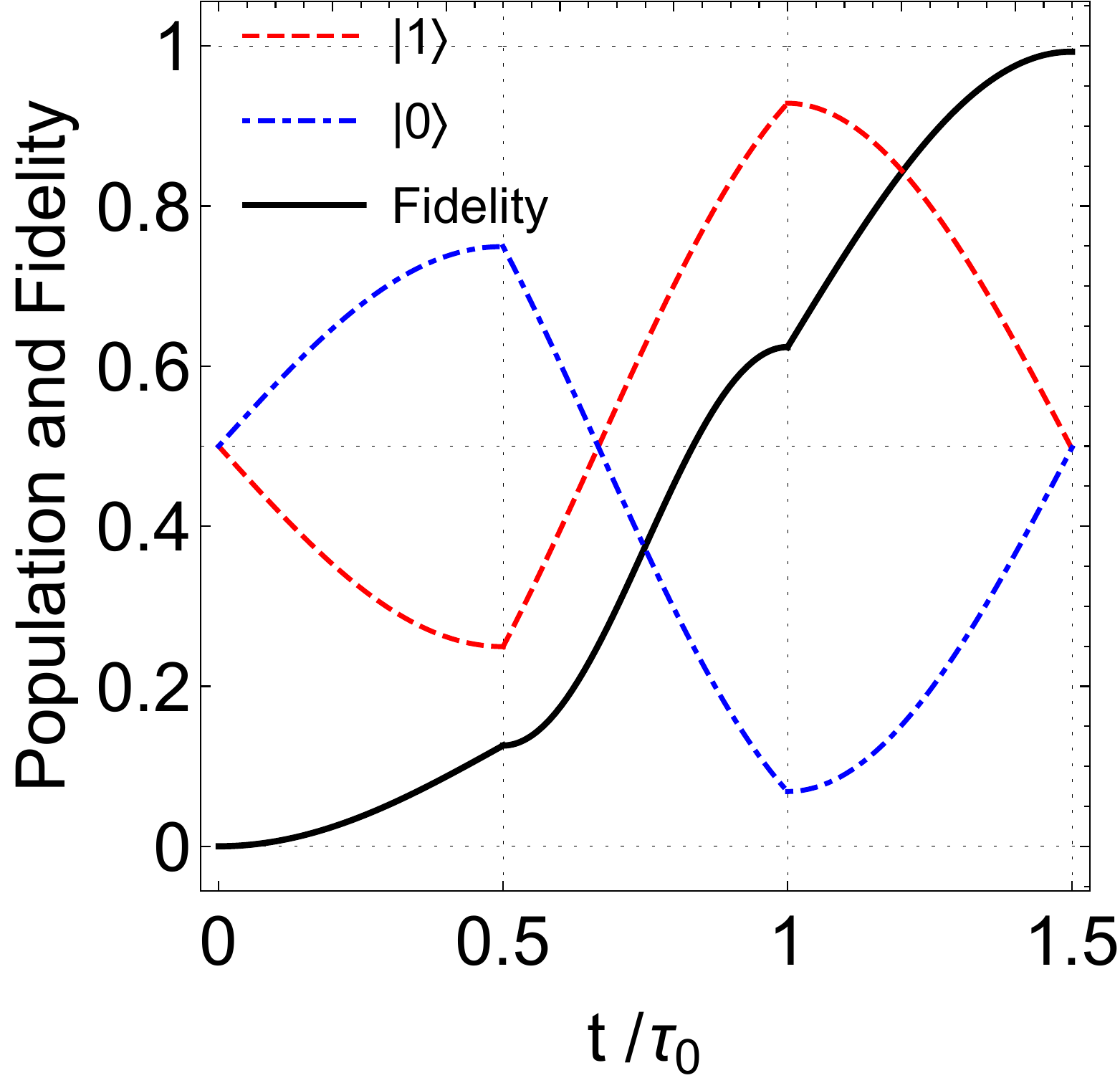}
\end{center}
\caption{(Left) The unconventional triangular path for the realization of $U_{Z}(\tau)$ gate;
(Right) The corresponding evolutions of state populations ($|1\rangle$: red dashed line; $|0\rangle$:
blue dot-dashed line) and the fidelities (black solid line) along three separate paths.}
\label{fig4}
\end{figure}

For example, we can choose the path in this way (see Fig.\ref{fig4}): the parameters $[\theta(t),\varphi(t)]$ start from
the north pole $(0,\varphi_{0})$ to the point $(\frac{\pi}{2},\varphi_{0})$ along the great circle $\varphi(t)=\varphi_{0}$,
then evolve along the equator to $(\frac{\pi}{2},\varphi_{0}+\frac{\pi}{2})$, and finally return back to the north pole along
the great circle $\varphi_{0}+\frac{\pi}{2}$. Along this path, we set $\eta=1$ to make Rabi frequency $\Omega(t)$ and
detuning $\Delta(t)$ both reaching maximum at the same time. Therefore, the Rabi frequencies of the control pulses are
\begin{equation}
\label{Rabi}
\Omega_{R}(t)e^{-i\phi(t)}=
\begin{cases}
\dot{\theta}(t)e^{-i(\varphi_{0}-\frac{\pi}{2})},\quad 0\leq t\leq\tau_{1},\\
\dot{\varphi}(t) e^{-i\varphi(t)},\quad\tau_{1}< t\leq\tau_{2},\\
\dot{\theta}(t)e^{-i\varphi_{0}},\quad\tau_{2}<t\leq\tau,
\end{cases}
\end{equation}
and their respective pulse areas and detunings satisfy
\begin{eqnarray}
\label{Rabi3}
\int_{0}^{\tau_{1}}\Omega_{R}(t)dt &=&\frac{\pi}{2},\,\Delta(t)=0,\notag\\
\int_{\tau_{1}}^{\tau_{2}}\Omega_{R}(t)dt &=&\frac{\pi}{2},\,\Delta(t)=\dot{\varphi}(t), \\
\int_{\tau_{2}}^{\tau}\Omega_{R}(t)dt &=&-\frac{\pi}{2},\,\Delta(t)=0.\notag
\end{eqnarray}
The total operation time along this path by the square-wave pulses is
$\frac{1}{2}\tau_0+\frac{1}{2}\tau_0+\frac{1}{2}\tau_0 = 1.5\tau_{0}$, which clearly demonstrates
that the unconventional GQC owns the shortest gate time comparing with the conventional ones. In addition,
although the conventional geometric gate driven by Eq.(\ref{Rabi2}) accumulates only
pure geometric phase and the unconventional geometric gate driven by Eq.(\ref{Rabi3}) involves nonzero
dynamical phase, both target gates acquire the same total phase that is only dependent on the global
geometric features. Accordingly, the Hamiltonians for convectional and unconventional gates share
similarities in the designed forms. More importantly, the freely adjusting parameter $\eta$ in our
theory can propose more optimal strategies to do GQC under relaxing experimental conditions.

\subsection{Two-qubit gate}

Moreover, a conditional two-qubit gate will be realized if we use two different pairs of orthogonal
cyclic states of the target qubit, conditioned on the state of another control qubit. The target qubit
is exploited by the electron spin of NV center and one nearby $^{13}\mathrm{C}$ nuclear spin as the
control qubit. In this case, a product one-qubit basis
$\{|0\rangle,|1\rangle\}\otimes\{|\downarrow\rangle, |\uparrow\rangle\}$ serves as a two-qubit computational
basis, which are coupled by different state-selective pulses and radio-frequency fields \cite{Liang,Two1}.
Under the parametric controls of the pulses, the effective Hamiltonian of this two-qubit system has an extensible form
\begin{equation}%
\hat{H}_{2}=\hat{H}_{\uparrow}\oplus\hat{H}_{\downarrow},
\end{equation}
that is
\begin{eqnarray*}
H_{2}=
 \frac{1}{2}\left[
  \begin{array}{cccc}
  \Delta_{\uparrow}(t) & \Omega(t) e^{-i\phi(t)} & 0 & 0\\
  \Omega(t) e^{i\phi(t)} & -\Delta_{\uparrow}(t) & 0 & 0\\
  0 & 0 & \Delta_{\downarrow}(t) & \Omega(t) e^{-i\phi(t)}\\
  0 & 0 &  \Omega(t) e^{i\phi(t)} & -\Delta_{\downarrow}(t)\\
   \end{array}
\right]
\end{eqnarray*}
in the two-qubit basis. In their respective subspaces of $\{\left\vert 1,\uparrow\right\rangle,|0,\uparrow\rangle\}$
and $\{|1,\uparrow\rangle,|0,\downarrow\rangle\}$, $H_{\uparrow}$ and $H_{\downarrow}$ can selectively satisfy
Eq.(\ref{U0t}) if the pulse frequency is on resonance with one computational state $|1,\uparrow\rangle$ and
$|0,\uparrow\rangle$, and far detuned from another computational state $|1,\downarrow\rangle$ and  $|0,\downarrow\rangle$
with a detuning $\delta\omega=\Delta_{\uparrow}-\Delta_{\downarrow}$.
The unwanted mixing effect caused by the coupling with the subspace of nuclear spin pointing downward can
neglect when $\Delta_{\downarrow}<\delta\omega$ is satisfied for
geometric gate \cite{Duan,Two1,Two2}. With the same routine design as that of a single-qubit gate $\hat{U}_{sq}$,
we can achieve the nontrivial geometric two-qubit gate as
\begin{equation}%
\hat{U}_{tq}=|\uparrow\rangle\langle\uparrow|\otimes \hat{U}_{sq}+|\downarrow\rangle\langle\downarrow|\otimes I.
\end{equation}

\section{performance of unconventional gates}

Now we check the reliability of the quantum geometric gates designed by our method in an open system. The performance of a $U_Z$ gate, in this case, can be simulated by using the Lindblad master equation as
\begin{equation}%
\dot{\rho}(t)=i\left[\rho(t),\hat{H}(t)\right]+\frac{1}{2}\left[\gamma_{1}\mathcal{L}(\hat{\sigma}^{+})%
+\gamma_{2}\mathcal{L}(\hat{\sigma}_{z})\right],
\end{equation}
where $\rho(t)$ is the density matrix of the designed system, and
$\mathcal{L}(\hat{A})=2\hat{A}\rho\hat{A}^{\dag}-\hat{A}^{\dag}\hat{A}\rho-\rho\hat{A}^{\dag}\hat{A}$ is the Lindbladian of
operator $\hat{A}$ ($\hat{\sigma}^{+}\equiv |1\rangle\langle 0|, \hat{\sigma}_{z}\equiv|1\rangle\langle 1|-|0\rangle\langle 0|$).
The decoherent effects of the environment are considered by the damping rates $\gamma_{1}$ and $\gamma_{2}$, respectively.
In our simulations, the decay and dephasing rates are set $\gamma_{1}=\gamma_{2}=4\times10^{4}$ Hz \cite{Duan}.
Suppose that the qubit is initially prepared in the state $ |\Psi(0)\rangle=\frac{1}{\sqrt{2}}(|0\rangle+|1\rangle)$
and the target state is $|\Psi_{Z}\rangle$, the time-dependence of the
state populations and the state fidelities $F(\tau)=|\langle\Psi_{Z}|\Psi(\tau)\rangle|^{2}$ of the $U_{Z}$
gates in different designed paths are shown in Figs.\ref{fig2}, \ref{fig3} and \ref{fig4}, respectively.
We can obtain a final fidelities of $99.35\%$ for the conventional geometric gate with an ``orange-slice" path,
$99.57\%$ for the conventional geometric gate in the triangular path, and $99.67\%$ for the unconventional geometric gate.

\section{conclusion}

In conclusion, we proposed a design method to implement conventional and unconventional nonadiabatic GQC
under the framework of SU(2) transformation. Our protocol relaxes the constraints imposed for the driving
Hamiltonian in the scheme of NHQC, and we can use the inversely designed Hamiltonian to realize the nonadiabatic
geometric gate with a desired evolutionary path. Our scheme can minimize the operation time needed for high-fidelity
geometric gates and is better combined with experimental techniques on a general computational platform. In other
words, our schemes have more selectivity than the previous ones towards possible designs of optimal protocols. To
show its potential applications, we simulate the performances of the geometric gates along different parametric
paths in the NV center system, and shorter evolution times and higher gate fidelities than the previous schemes
are explicitly confirmed.

\section*{ACKNOWLEDGMENTS}
We thank Bao-Jie Liu for fruitful discussions.
This work was supported by the National Natural Science Foundation of China for
emergency management project (Grants No.11447025 and No.11847308).

\end{document}